\documentclass[usenatbib,usegraphicx]{mn2e}

\usepackage{psfig}

\newcommand{\HI}{\mbox{H\,{\sc i}}}
\newcommand{\Lya}{Ly-$\alpha$}
\newcommand{\Lyb}{Ly-$\beta$}
\newcommand{\Lyg}{Ly-$\gamma$}
\newcommand{\CIV}{\mbox{C\,{\sc iv}}}
\newcommand{\CIII}{\mbox{C\,{\sc iii}}}
\newcommand{\CII}{\mbox{C\,{\sc ii}}}
\newcommand{\NV}{\mbox{N\,{\sc v}}}
\newcommand{\NIII}{\mbox{N\,{\sc iii}}}
\newcommand{\NII}{\mbox{N\,{\sc ii}}}
\newcommand{\OVI}{\mbox{O\,{\sc vi}}}
\newcommand{\MgII}{\mbox{Mg\,{\sc ii}}}
\newcommand{\AlII}{\mbox{Al\,{\sc ii}}}
\newcommand{\SiIV}{\mbox{Si\,{\sc iv}}}
\newcommand{\SiIII}{\mbox{Si\,{\sc iii}}}
\newcommand{\SiII}{\mbox{Si\,{\sc ii}}}
\newcommand{\SIII}{\mbox{S\,{\sc iii}}}
\newcommand{\SIV}{\mbox{S\,{\sc iv}}}
\newcommand{\FeIII}{\mbox{Fe\,{\sc iii}}}
\newcommand{\FeII}{\mbox{Fe\,{\sc ii}}}

\newcommand{\kms}{km s$^{-1}$}
\newcommand{\zem}{$z_{\rm em}$}
\newcommand{\zabs}{$z_{\rm abs}$}

\title[Chemical abundances in QSO host galaxies and
  environments]{Chemical abundances in QSO host galaxies and
  environments from narrow absorption line systems\thanks{Based on 
   material collected with the European Southern
   Observatory Very Large Telescope operated on Cerro Paranal
          (Chile). Proposal 116.A-0106, 65.O-0299, 67.A-0078 and
  69.A-0204}}  
\author[V. D'Odorico, S. Cristiani, D. Romano, G.L. Granato \&
    L. Danese]{Valentina 
    D'Odorico,$^{1}$\thanks{E-mail: 
    dodorico@sissa.it} Stefano Cristiani,$^{2}$ Donatella  Romano,$^{3}$ 
    \newauthor Gian Luigi Granato$^{4}$ and Luigi Danese$^{1}$ \\
    $^{1}$SISSA/ISAS, via Beirut 2, I-34014 Trieste, Italy \\
    $^{2}$INAF - Osservatorio Astronomico di Trieste, 
               via  G.B. Tiepolo, 11, I-34131 Trieste, Italy \\
    $^{3}$INAF - Osservatorio Astronomico di Bologna,
               via Ranzani 1, I-40127 Bologna, Italy \\
    $^{4}$INAF - Osservatorio Astronomico di Padova, 
              vicolo dell'Osservatorio 5, I-35122 Padova, Italy} 

\begin{document}

\date{Received; accepted}

\pagerange{\pageref{firstpage}--\pageref{lastpage}} \pubyear{2004}

\maketitle

\label{firstpage}

\begin{abstract}
We determined C, N and $\alpha$-element relative abundances in the gas
surrounding six QSOs at an average redshift of $< z > \simeq 2.4$, by
studying six narrow associated absorption systems in UVES
high-resolution spectra.   
We found five systems with a metallicity (measured by C/H) consistent
or above the solar value. 
The ionization structure observed in the associated systems is
clearly different from that of the intervening ones, indicating that
the associated systems are influenced by the strong UV flux from the
QSO. 
There is a possible correlation (anticorrelation) between [N/C]
([Si/C]) and [C/H]  of the studied associated systems, 
and [N/C]~$\ge 0$ when [C/H]~$\ge 0$.  
We have compared  these observational results with the predictions of a
model simulating the joint evolution of QSOs and their spheroidal
hosts. The agreement turns out to be very good, in particular, the
case envisaging massive haloes and high star-formation rates
recovers both the correlation between [N/C] and [C/H] and the
anticorrelation for [Si/C] vs. [C/H].   
Narrow associated absorption systems prove to be powerful tracers of the
chemical abundances in gas belonging to high redshift spheroidal galaxies. 
The outflow of this same gas, triggered by the QSO feedback, is
probably going to contribute to the early enrichment of the
surrounding intergalactic medium.  
A larger statistics, possibly increasing the number of ionisation
stages, chemical elements and the redshift range, would allow us to
put firm constraints on detailed chemical evolution models of galaxies
at high redshifts.  
\end{abstract}


\section{Introduction}

In this work, we want to address the star formation history and the 
evolution of massive early-type galaxies at high redshifts by
measuring in a reliable way the metallicity and the chemical
abundances of gas belonging to host galaxies and environments of QSOs.  

Once considered rare and exotic objects, QSOs could instead  
represent a necessary phase in the evolution of massive early-type 
galaxies. 
This interpretation is supported by several pieces of evidence:  
Massive Dark Objects (MDOs, generally interpreted as dormant black
holes) with masses in the range $\sim 10^6$~-~$3\times10^9$ M$_{\sun}$
are present in essentially all local galaxies with a substantial
spheroidal component \citep[see ][ for a review]{korm:geb}, on the
other hand the host galaxies of low redshift powerful AGN (radio-loud
and radio-quiet QSOs and radio galaxies) are, in all the studied
cases, luminous elliptical galaxies with $L>L^{\ast}$ \citep{dunlop03}. 

The observational properties inferred for cluster and field elliptical
galaxies up to redshift $z \sim 1$ imply a high uniformity and
synchronization in the galaxy formation process
\citep[e.g.][]{ellis97,bernardi98}.  The evolution with
redshift of their optical-IR colours \citep{stanford98} is consistent
with the passive evolution of an old stellar population formed at $z
\ge 2-3$ and the measured positive [Mg/Fe] elemental ratio 
can be explained by a short and intense star formation burst
\citep[e.g. ][]{wortheyeal,matteucci94}. 

In the standard framework of the hierarchical evolution of 
structures in a cold dark matter (CDM) universe large objects
form by a sequence of mergers of smaller proto-galaxies. 
In particular, massive ellipticals are generated at low redshifts ($z
\le 2$) from the merger of two large disk galaxies which formed stars
at a constant moderate rate up to that moment
\citep[e.g. ][]{baugheal,KC98}.  
In the merging event, the black holes (BHs) pre-existing in the
progenitor galaxies coalesce and a fraction of the cold gas is
accreted by the new BH which activates as a QSO, the rest of the cold
gas is transformed into stars in a sudden burst
\citep{WB98,kauff:haeh,marta03,mencieal03}. 

A different prescription in the framework of the hierarchical scenario
is the {\sl anti-hierarchical baryon collapse} where the formation of
stars and of the central BH takes place on shorter time-scales within more
massive dark matter haloes \citep*{monaco00,granato01,archi02,granato04}. 
Supernova heating and QSO feedback are the physical processes that
reverse the order of formation of galaxies compared to that of DM
haloes because they slow down star formation most effectively in
shallow potential wells.  
In the more massive  DM haloes star formation goes on rapidly causing at the
same time the growth of the central BH which accretes the cold gas slowed
down by the radiation drag. When the QSO activates, strong winds
originate sweeping the interstellar medium and halting both the star
formation and the BH growth.   
The time delay between the star formation onset and the peak of the QSO
activity is again shorter for larger haloes. For the most massive galaxies
($M_{\rm halo} \ga  10^{12}\ M_{\sun}$) virializing at $3 \le z \le 6$,
this time is $< 1$ Gyr, implying that the bulk of star formation may be
completed before type Ia supernovae have the time to significantly enrich
the interstellar medium with iron. 
A detailed analysis of the chemical evolution expected for this model is
reported in  \citet{romano02}. 

The two above  described scenarii predict different chemical
abundances, in particular at redshifts larger than $\sim 2$. 
The metallicity and the elemental abundances of high redshift galaxies
are hard to measure; on the other hand, high and intermediate resolution
spectra of QSOs at redshifts as large as $z \sim 6$ can be easily
obtained with the present instrumentation.

We studied associated narrow\footnote{The adjective ``narrow'' is used
  to distinguish this class of absorptions from the Broad
  Absorption Lines characterised by FWHM~$> 2000$ \kms\ and arising in
  gas ejected by the QSO at large velocities (see also Sections~2 and
  7)} absorption lines exploiting high resolution,  
high signal-to-noise ratio spectra of $2 < z< 3$ QSOs obtained with the 
UVES spectrograph and a model for the
photoionisation of the gas to derive chemical abundances in the QSO
environments. Our results suggest that at these redshifts the gas
associated with the QSO and with its host galaxy has already been
enriched by the products of an intense star formation episode. 
 
Section~2 introduces  the diagnostics that we used to determine the  
chemical abundances in QSO environments and reports previous results.    
In Section~3 we describe the selection criteria and the characteristics  
of our sample of associated narrow absorption line systems; the
adopted photoionisation model and the methodology are reported in
Section 4. 
Section~5 is devoted to the description of our results, which are
compared with model predictions in Section~6. In Section~7, we
summarise the results on QSO chemical abundances obtained using other
methods. 
We draw our conclusions in Section~8.

\section{Chemical abundances in the vicinity of QSOs measured with
  associated narrow absorption lines}
 
The ``narrow'' absorption lines (NALs) are the most numerous
in QSO spectra.  
A practical way to define them is that they have to be narrow enough
that important UV doublets are not blended, i.e. to have FWHM~$<200$ to 300
\kms.   

Here we are concerned only with the NALs falling within $\pm 5000$ \kms\ 
of the systemic redshift: the so called ``associated'' systems (AALs). 
In particular, since we are interested in determining the metal abundances
and the physical conditions in the gaseous environments close to QSOs,
we would like to identify which of these systems are also ``intrinsic'',
that is physically associated with the QSO.   
The intrinsic nature of individual AALs can be inferred from various
indicators, for example (1) time variability in the absorption lines
requiring dense,  
compact regions and thus intense radiation fields near the QSO for 
photoionisation, (2) high space densities measured directly from 
excited-state fine-structure lines, (3) partial coverage of the emission 
source measured via resolved, optically thick lines with too-shallow 
absorption troughs, (4) spectropolarimetry that reveals an unabsorbed or 
less absorbed spectrum in polarized light, (5) smooth and relatively 
broad absorption line profiles that are unlike intervening absorbers, and  
(6) higher ionisation states than intervening absorbers. 

Previous studies on AALs have found that in general these systems have
solar or supersolar metallicities
\citep*[e.g. ][]{wamplereal93,mollereal94,ppjeal94, 
  hamanneal97,ppj:srian,sria:ppj00}. 
In a few cases moderately subsolar metallicities are observed, but they 
are still significantly higher than those found in intervening systems at the
same redshifts \citep{savaglioeal97}.  
Besides, a marked change in the metallicity of QSO absorption systems
from values smaller than 1/10 solar to solar or larger values is
observed at a blueshift of $\sim 15000$ \kms\ relative to the QSO
emission lines  \citep{ppjeal94,fra:gratton}.

In order to derive reliable metallicities and chemical abundances we had to
carefully select our AALs. The adopted criteria and the sample are
described in the following section. 

\section{The sample}

\subsection{Selection criteria and data analysis}

We have considered the 16 QSO spectra of the ESO Large Programme  
``The Cosmic Evolution of the Intergalactic Medium'' (116.A-0106A, 
P.I. J. Bergeron) available to the public and the 6 QSO spectra of our pair
archive described in \citet*{vale02}.  
All of them were obtained with the UVES spectrograph at the Kueyen unit of
the ESO VLT (Cerro Paranal, Chile) with a resolution $R \sim 40000$ and a
wavelength coverage $310 < \lambda < 1000$ nm. The signal to noise ratio
(S/N) of the Large Programme spectra is S/N~$\sim 50$, while in our spectra
it is varying between S/N~$\sim 5$ and $20$ depending on the brightness of
the observed QSO. 
We used the Large Programme spectra reduced with the automatised, refined
version of the UVES pipeline \citep{ball00} devised by Bastien Aracil
\citep[see ][]{aracil03}. 
The normalisation has been carried out by manually selecting spectral
regions not affected by evident absorption and by interpolating them with a
spline function of 3rd degree. 
 
In these spectra we looked for \CIV\ and \NV\ doublets within $\pm
5000$ \kms\ from the emission redshift of the QSO.  If both
absorptions were detected at the same redshift we considered the system as
a candidate intrinsic absorber and we looked for other associated ions
(like \CII, \CIII, \NIII, \SiII, \SiIII, \SiIV, \MgII, etc).  

Among the 22 QSOs in the sample, 16 showed associated \CIV\
doublets summing up to a total of 34 systems of which 15 had
detectable \NV\ lines too.  
In 11 of the 19 systems without \NV, the rest equivalent width of the 
\CIV\ $\lambda 1548$ absorption line was lower than $\sim 0.03$
\AA. Since in general the \NV/\CIV\ equivalent width ratio is lower
than one, the \NV\ transition even if present could be under our
detection limit. 
Four of the 8 systems with stronger \CIV\ doublets, have \CIII/\CIV\
equivalent width ratios larger than one. As we will discuss  
in Section~5 this is a characteristic of intervening systems while the 
contrary is true for associated ones. 
The remaining 4 systems do not show hints of intrinsic nature by any
of the six properties listed in the previous section, 
except one for which an effect of partial coverage is detected. 

In order to constrain the photoionisation model (see Section~4) and to
get reliable results for the metallicity and the element relative
abundances we required also that at least two ionisation states of the
same chemical element were detected and that the corresponding
absorption lines were not saturated.   
This requirement reduced the number of viable systems to 6. 

The systems excluded from the sample by our defining criteria can be 
divided into two groups: the first one is formed by those systems for which 
we did not observe enough ionic transitions to constrain the photoionisation 
model, for the majority of them only a weak  \CIV\ doublet was detected. 
In some cases, more lines could be observed extending the spectral 
wavelength coverage with new observations. 
The second group is made by a few strong systems where most of the lines 
are saturated not allowing the determination of reliable column densities.

\begin{table}
\caption{Atomic parameters used in the fitting of the lines which differ
  from \citet{morton}}
\label{atom}
\begin{tabular}{@{}lll}
\hline
Ion & Rest wavel. & Osc. strength \\
& (\AA) & \\
\hline
\CIV$\ldots$ & 1548.204(2)$^1$ & 0.1908 \\
            & 1550.781(2)$^1$ & 0.09522 \\
\MgII$\ldots$ & 2796.3543(2)$^2$ & 0.6123 \\
             & 2803.5315(2)$^2$ & 0.3050 \\
\AlII$\ldots$ & 1670.7886(1)$^1$ & 1.833 \\
\SiII$\ldots$ & 1304.3702 & 0.086$^3$ \\
              & 1526.70698(2)$^1$ & 0.110$^3$ \\
              & 1808.01288(1)$^1$ & 0.0022$^4$ \\
\SiIV$\ldots$ & 1393.76018(4)$^1$ & 0.5140 \\
             & 1402.77291(4)$^1$ & 0.2553 \\
\hline
\end{tabular}

\medskip
$^1$ Griesmann \& Kling (2000); 
$^2$  Pickering et al. (2000; 2002); 
$^3$  Spitzer \& Fitzpatrick (1993); 
$^4$  Bergeson \& Lawler (1993)
\end{table}

We fitted the observed absorption lines with Voigt profiles in the LYMAN
context of the MIDAS reduction package \citep{font:ball}. Adopted atomic 
parameters differing from those reported in \citet{morton} are given in  
Table~\ref{atom}. 

\vskip 12pt

In the following we will describe in detail the selected systems. The
properties of the background QSOs are reported in Table~\ref{qsos}.  

\begin{table}
\caption{Relevant parameters of the QSOs studied in this work. The
  absolute magnitudes are taken from the 10th edition of the QSO catalog
  by V\'eron-Cetty \& V\'eron (2001)}
\label{qsos}
\begin{tabular}{lllc}
\hline
QSO Name & \zem & Magnitude & Abs. Mag.\\
\hline
UM680 & 2.1439 & V=18.6 & -27.5 \\
UM681 & 2.1219 & B$_J$ = 18.8 & -27.3 \\
HE1158-1843 & 2.453 & V=16.93 & -29.4 \\
Q2343+1232 & 2.549 & V=17.0 & -29.7\\
Q0453-423 & 2.661 & V=17.06 &-29.2  \\
PKS0329-255 & 2.685 & V=17.51 & -29.2  \\ 
\hline 
\end{tabular}
\end{table}

\subsection{Individual systems}

\begin{figure}
\includegraphics[width=8cm,height=9cm]{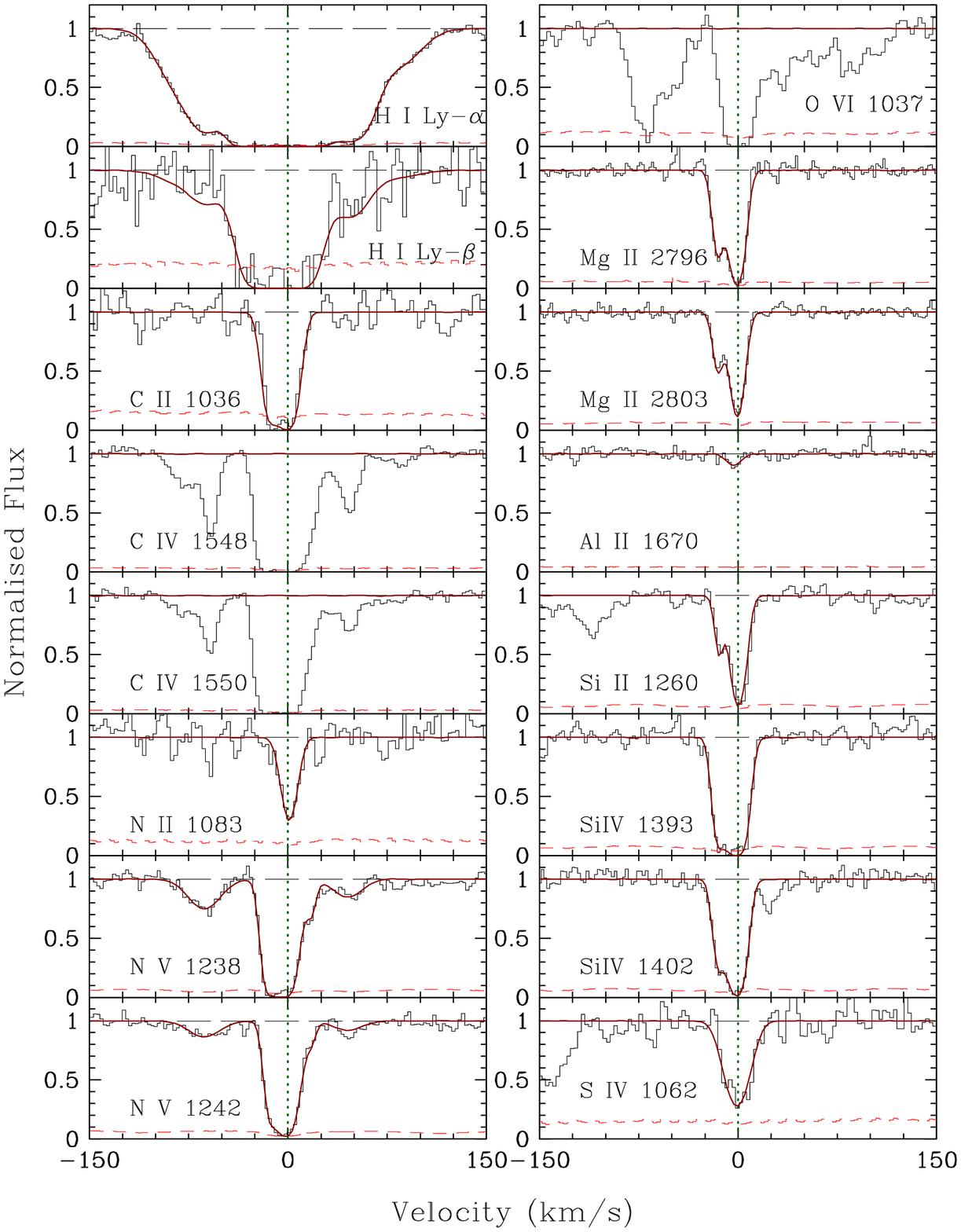}
\caption{Ionic transition lines observed at redshift $z=2.123$ (marked by
  the vertical dotted line) in the spectrum of QSO UM680. The result
  of the best fitting for the analysed absorptions is overplotted on
  the spectrum. The short-dashed
  line represents the noise}
\label{um680_z212}
\end{figure}

\begin{figure}
\includegraphics[width=8cm,height=7cm]{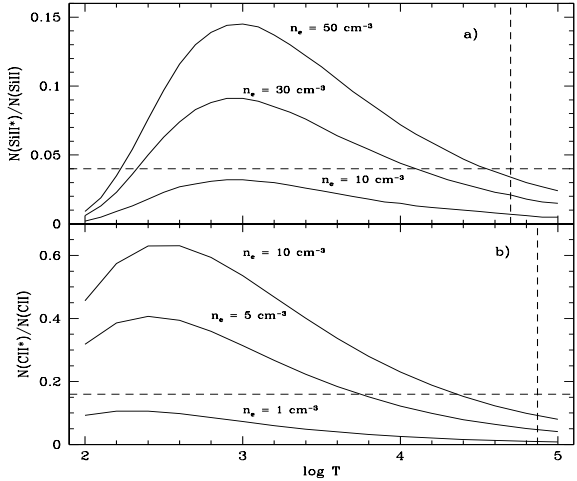}
\caption{{\bf a)} Theoretical \SiII$^{\ast}$(1264)/\SiII(1260) column
  density ratio as a 
  function of temperature at increasing electron density (from bottom to
  top, as marked on the picture). The dashed lines
  represent the upper limits on the \SiII$^{\ast}$/\SiII\ column density
  ratio (horizontal line) and on the temperature (vertical line) for the
  absorption system at \zabs~$=2.123$ in the spectrum of UM680; {\bf b)}
  Theoretical \CII$^{\ast}$(1335)/\CII(1334) column
  density ratio as a function of temperature. Here the dashed lines
  correspond to the limits obtained for the 
  absorption system at \zabs~$=2.122$ in the spectrum of UM681} 
\label{ne_c2si2} 
\end{figure}


\begin{table}
\caption{Parameters obtained fitting the
  absorption system at \zabs~$\sim 2.123$ in the spectrum of UM680}
\label{cd_um680}
\begin{tabular}{llcc}
\hline
Ion & Redshift & log N & b  \\
& & (cm$^{-2}$) & (\kms) \\
\hline 
\HI$\ldots$  & $2.122327\pm0.00005$ & $13.7\pm0.2$ & $30\pm3$ \\
             & $2.12247\pm0.00003$ & $13.2\pm0.3$ & $15\pm4$ \\
             & 2.12302(f)$^1$ & 15.2$^2$ & 20(f) \\
             & $2.123586\pm 0.000008$ & $13.92\pm0.03$ & $19\pm1$ \\
             & $2.12399\pm0.00004$ & $13.0\pm0.1$ & $21\pm4$ \\ 
\CII$\ldots$ & 2.12295(f) & $14\pm0.2$ & $4.5\pm0.6$ \\
             & 2.1231(f) & $14.6\pm0.2$ & $6.2\pm 0.6$ \\
\NII$\ldots$ & $2.123110\pm0.000009$ & $13.79\pm0.06$ & $6.2\pm0.6$ \\
\NV$\ldots$  & $2.122445\pm0.000008$ & $13.21\pm0.03$ & $16\pm1$ \\
             & $2.12293\pm0.00002$ & $13.7\pm0.1$ & $4\pm1$ \\
             & $2.123058\pm0.000008$ & $14.42\pm0.05$ & $9.1\pm0.5$ \\
             & $2.123284\pm0.000008$ & $12.77\pm0.07$ & 3(f) \\
             & $2.12357\pm0.00002$ & $12.93\pm0.07$ & $14\pm3$ \\
\MgII$\ldots$& $2.122941\pm0.000004$ & $12.51\pm0.02$ & $3.6\pm0.4$ \\
             & $2.123091\pm0.000004$ & $13.11\pm0.03$ & $5.2\pm0.3$ \\
\AlII$\ldots$& $2.12306\pm0.00002$ & $11.25\pm0.09$ & $6.2\pm0.6$ \\
\SiII$\ldots$& $2.122949\pm0.000005$ & $12.29\pm0.06$ & $2.5\pm0.9$ \\
             & $2.123101\pm0.000002$ & $13.03\pm0.02$ & $5.5\pm0.3$ \\
\SiIV$\ldots$& $2.122940\pm0.000007$ & $13.18\pm0.06$ & $4\pm0.7$ \\
             & $2.123091\pm0.000007$ & $13.9\pm0.1$ & $6.4\pm0.6$ \\
\SIV $\ldots$& $2.123091\pm0.000009$ & $14.39\pm0.05$ & $11\pm1$ \\
\hline
\end{tabular}

\medskip
$^1$ We indicate with an (f) those parameters that were assumed and
not determined by the fitting process \\
$^2$ For the uncertainty on this determination see the discussion in
the text
\end{table}

\centerline{\bf {System at \zabs~$= 2.123$  towards UM680}} 

\noindent
The system at \zabs~$=2.123$  (\zem~$-$~\zabs~$\sim 1990$ \kms) has strong 
\CIV, \NV, \OVI\ and \SiIV\ lines. Also \CII, \NII, \MgII, \AlII, \SiII\
and  \SIV\ are observed. The central components of \CIV\ are too heavily
saturated and for \OVI\  the signal-to-noise ratio is too low to determine
reliable column densities. 
Also the \HI\ \Lya\ line is saturated. In order to obtain an estimate
of the \HI\ column density for the velocity component corresponding to
the metal absorptions, we fixed its redshift as the average of the
two \SiII\ and \SiIV\ component redshifts. 
The velocity profiles of  the  \Lya\ and \Lyb\ transitions are not
constraining the fitting process which does not converge to a unique
result.  We had then to adopt a Doppler parameter for the central
component, while leaving the other component parameters free.  
For a typical value $b=20$ \kms\ observed for \Lya\ lines at redshift
around 2.1 \citep[see ][]{kim2001} a  $\log N($\HI$) = 15.2$ is derived. 
To estimate the uncertainty on this determination, we determined
the \HI\ column densities for two limit values of the Doppler
parameter: $b=15$ \kms\ (consistent with the lower limit observed in
the $b$-distribution of the Lyman forest) and $b=30$ \kms\ (the
maximum value allowed by the \Lyb\ velocity profile). They turned out
to be $\log N($\HI$) = 16$ and $\log N($\HI$) =14.8$ respectively. 
In the following computation of chemical abundance ratios  for this 
velocity component  we will adopt as the error on the \HI\ column 
density 0.6 dex to take into account this uncertainty on its
determination.  
 
All the measured column densities are reported in Table~\ref{cd_um680} 
and the ionic transitions are shown in Fig.~\ref{um680_z212}.     

\noindent
The fine-structure transition absorptions \CII$^{\ast}$ corresponding
to the strong absorption lines due to \CII\ $\lambda\,1036,\,1334$ are  
absent. Besides, we did not observe the \SiII$^{\ast}$ $\lambda\,1264$  
fine-structure line associated with the \SiII\ $\lambda\,1260$
absorption.  
The upper limits on the \SiII$^{\ast}$/\SiII\ column density ratio for
the two observed components are $0.19$ and $0.04$. The corresponding
upper limits on the temperature derived from the Doppler parameters
are $\log\, T \le 4$ and $\log\, T \le 4.7$ respectively.  
Following \citet{sria:ppj00} we determined an upper limit on the electron
density of $n_{\rm e} \sim 10-30$ cm$^{-3}$ applying the average values of
the previously reported limits to the formula of pure collisional
excitation \citep[][ and references therein, note that excitation by
  hydrogen atoms is unimportant]{fitz:spitz}:    

\begin{equation}
\label{density}
\frac{N({\rm X~II}^{\ast})}{N({\rm X~II})} = \frac{n_{\rm
  e}\,\gamma_{12}({\rm e})}{A_{21} + n_{\rm
  e}\,\gamma_{21}({\rm e})},
\end{equation}

\noindent
where X is either silicon or carbon, $\gamma$ are the collision excitation
and de-excitation rates, and $A_{21}$ is the radiative decay rate (see
Fig.~\ref{ne_c2si2}). Since the gas is optically thin it is reasonable to
assume $n_{\rm e} \approx n_{\rm H}$.  

\vskip 12pt

\begin{figure}
\includegraphics[width=8cm,height=9cm]{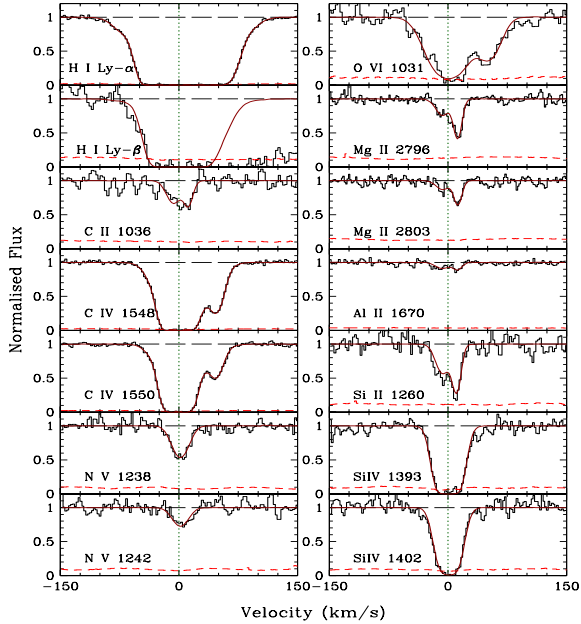}
\caption{Ionic transition lines observed at redshift $z=2.122$ (marked by
  the vertical dotted line) in the spectrum of QSO UM681. The result
  of the best fitting for the analysed absorptions is overplotted on
  the spectrum. The short-dashed
  line represents the noise}
\label{um681_z212}
\end{figure}

\centerline{\bf{System at \zabs~$=2.122$ towards UM681}}

\noindent
An absorption system at \zabs~$=2.122$ (\zem~$-$~\zabs~$\sim -10$ \kms) is
detected with lines due to \CIV, \NV, \OVI\ and \SiIV. The low ionisation
transitions of \CII, \NII, \SiII, \AlII\ and \MgII\ are present too (see
Fig.~\ref{um681_z212}).    
The \HI\ \Lya\ absorption is saturated and the \Lyb\ line  is
partially blended. 
To measure the \HI\ column density we fixed the redshift of the main
component as the average of the two components observed in the 
low-ionisation transitions. We used only the \Lya\ line to determine
the fitting parameters and verified the consistency of the result with
the blue wing of the \Lyb\ velocity profile which looks free from
major blending.  
All the measured column densities are reported in
Table~\ref{cd_um681}.  
  
\begin{table}
\caption{Parameters obtained fitting the
  absorption system at \zabs~$\sim 2.122$ in the spectrum of UM681 }
\label{cd_um681}
\begin{tabular}{llcc}
\hline
Ion & Redshift & log N & b  \\
& & (cm$^{-2}$) & (\kms) \\
\hline 
\HI$\ldots$  &$2.1214\pm0.0001$ & $13\pm0.3$ & $29\pm7$ \\
             &2.122015(f) & $15.7\pm0.2$ & $26\pm1$ \\
             &$2.122500\pm0.000008$ & $14.16\pm0.02$ & $22\pm1$ \\
             &$2.1229\pm0.0001$ & $12.4\pm0.4$ & $17\pm9$ \\
\CII$\ldots$ &$2.121930\pm 0.000003$ & $13.29 \pm 0.1$  & $11 \pm 2$   \\
             &$ 2.122113\pm0.000003$ & $13.41\pm 0.07$ & $6\pm 1$ \\
\CIV$\ldots$ &$2.12183\pm0.00002$ & $13.6\pm 0.2$  & $21 \pm 3$  \\
             &2.122016(f) & $15.23\pm 0.04$  & 13.5(f)  \\
             &$2.12235\pm 0.00002$ & $13.62 \pm 0.06$  & $23 \pm 4$ \\
             &$2.122520\pm0.000002$& $13.3 \pm 0.1$  & $9 \pm 1$  \\
\NV$\ldots$  &$2.1220193\pm0.000004$ & $13.49 \pm 0.03$  & $13 \pm 1$   \\
\OVI$\ldots$ &$2.122006\pm0.000009$ & $14.5 \pm 0.03$  & $28 \pm 2$  \\
             &$2.12254\pm0.00002$ & $13.92\pm0.06$ & $18\pm0.2$ \\
\MgII$\ldots$&$2.121932\pm0.000008$ & $12.1 \pm 0.1$ & $11 \pm 2$  \\
             &$2.122128\pm 0.000003$ & $12.38\pm 0.07$ & $6\pm 1$  \\
\AlII$\ldots$&2.121932(f)  & $11.40\pm0.07$ & 11(f) \\
             &2.122128(f)  & $11.28 \pm0.08$ & 6(f) \\
\SiII$\ldots$ & $2.121930\pm 0.000008$& $12.54 \pm 0.09$  & $11 \pm 2$  \\
              & $2.122113\pm 0.000002$& $12.68\pm 0.07$ & $6\pm 1$ \\
\SiIV$\ldots$ &$2.122019\pm0.000001$ & $14.12 \pm 0.05$  & $14.4 \pm 0.5$ \\
\hline
\end{tabular}
\end{table}

\noindent
Also for this absorption system we did not detect the fine-structure
transition lines of \CII\ and \SiII. 
We obtained the upper limits on the column density ratio \CII$^{\ast}$
$\lambda\, 1335$ over \CII\ $\lambda\, 1334$ and on the gas
temperature for the two components of the system:
\CII$^{\ast}$/\CII~$\la 0.16$ and $\log\, T < 4.87$, and
\CII$^{\ast}$/\CII~$\la 0.18$ and $\log\, T < 4.66$ respectively.  
In Fig.~\ref{ne_c2si2}, we plot the curves of the \CII\ ratio as a
function of temperature for constant $n_{\rm e}$ as given by
eq.~\ref{density}. 
The observed limits are consistent with an electron density of $n_{\rm e}
\le 1-5$ cm$^{-3}$. 

\noindent
Note that this absorption system and the one along the line of sight to
UM680 are less than $\sim 100$ \kms\ apart in redshift  and that the two
lines of sight are separated by $\sim 700$ kpc proper distance at this
redshift.

\vskip 12pt

\begin{figure}
\includegraphics[width=8cm,height=8cm]{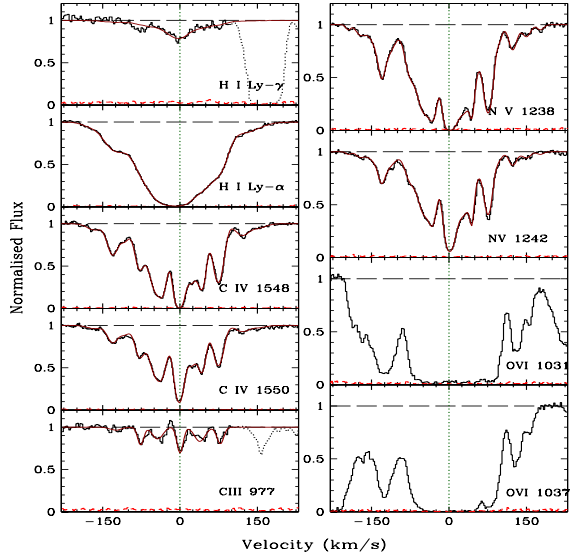}
\caption{Ionic transitions lines observed at redshift $z\sim 2.44264$
  (marked by the vertical dotted line) in the spectrum of QSO
  HE1158-1843. The zero level of the \CIV\ and \NV\ doublets has been
  modified to coincide with the minimum flux of the stronger line of
  the doublet in order to carry out the fitting. The same was done for
  the \Lya\ absorption line. The result of the best fitting for the
  analysed absorptions is overplotted on the spectrum. The
  short-dashed line represents the noise} 
\label{he1158_z244}
\end{figure}

\begin{figure}
\includegraphics[width=8cm,height=8cm]{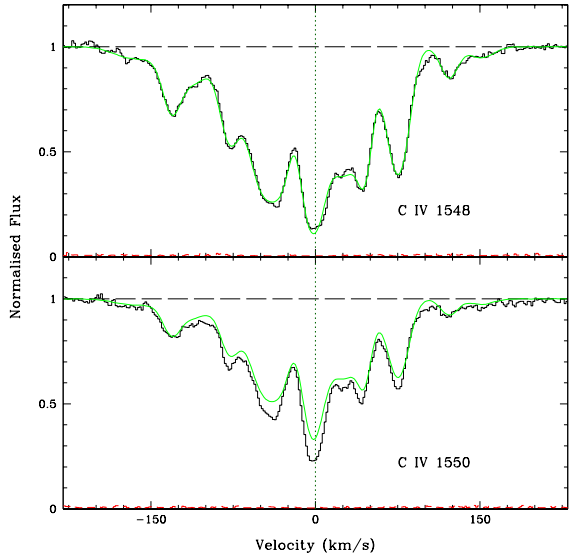}
\caption{\CIV\ doublet observed at $z\sim 2.4426$ in the spectrum of
  QSO HE1158-1843. The thick solid line represents the profile
  fitting based on the intensity of the \CIV\ $\lambda\,1548$ \AA\
  absorption only. When applied to the companion line of the doublet it 
  clearly shows the effect of partial coverage }
\label{part_cov}
\end{figure}

\begin{table}
\caption{Parameters obtained fitting the absorption system at
  \zabs~$\sim 2.4426$ in the spectrum of HE1158-1843} 
\label{cd_he1158}
\begin{tabular}{llcc}
\hline
Ion & Redshift & log N & b  \\
& & (cm$^{-2}$) & (\kms) \\
\hline
\HI$\ldots$   & $2.44061\pm0.00002$ & $12.02\pm0.05$ &   20 (f) \\   
              & $2.44195\pm0.00002$ & $13.72\pm0.04$ &  $33\pm2$ \\  
              & 2.442263 (f) & $13.65\pm0.05$ &  $23.2\pm0.8$ \\  
              & 2.442616 (f) & $14.22\pm0.02$ & $29.4\pm0.3$ \\  
              & 2.443143 (f) & $13.56\pm0.01$ & $22.0\pm0.3$ \\  
              & 2.443506 (f) & $13.38\pm0.01$ & $21.8\pm0.1$ \\  
              & 2.444044 (f) & $12.87\pm0.01$ & $41.0\pm0.9$ \\  
              & $2.441214\pm0.000008$ & $13.22\pm0.02$ &  $31\pm1$ \\  
\CIII$\ldots$ & $2.44121\pm0.00004$ & $11.70\pm0.09$ & $15\pm3$ \\
              & $2.441768\pm0.000004$ & $12.24\pm0.02$ & $6.9\pm0.6$\\
              & $2.44214\pm0.00001$ & $12.28\pm0.02$ & $13.2\pm0.9$ \\
              & $2.442643\pm0.000005$ & $12.43\pm0.02$ &  $6.8\pm0.5$ \\
              & $2.44308\pm0.00001$ & $12.38\pm0.02$ &  $13.7\pm0.9$\\
              & $2.443539\pm0.000004$ & $12.26\pm0.02$ & $8.2\pm0.4$\\
\CIV$\ldots$  & $2.44075\pm0.00001$ & $12.55\pm0.03$ & $20\pm1$ \\
              & $2.441137\pm0.000006$ & $13.12\pm0.02$ & $12.6\pm0.4$\\
              & $2.441411\pm0.000006$ & $12.90\pm0.04$ & $15\pm1$ \\
              & $2.441741\pm0.000001$ & $13.31\pm0.01$ & $10.6\pm0.3$\\
              & $2.44211\pm0.000006$ & $13.80\pm0.03$ & $17.2\pm0.6$\\
              & $2.44226\pm0.000006$ & $13.44\pm0.06$ & $11.2\pm0.4$\\
              & $2.442627\pm0.000006$ & $14.05\pm0.05$ & $10.7\pm0.1$\\ 
              & $2.442898\pm0.000006$ & $13.48\pm0.03$ & $10.5\pm0.6$ \\
              & $2.443131\pm0.000006$ & $13.57\pm0.01$ & $9.7\pm0.2$\\
              & $2.443499\pm0.000001$ & $13.56\pm0.03$ & $12.3\pm0.1$ \\
              & $2.444045\pm0.000002$ & $12.80\pm0.01$ & $17.7\pm0.6$ \\
              & 2.444378 (f) & $12.15\pm0.02$ & 10 (f)\\
\NV$\ldots$   & 2.440400 (f) & $12.05\pm0.04$ &  5  (f) \\
              & $2.440667\pm0.000008$ & $12.58\pm0.03$ & $11\pm 1$\\
              & $2.441144\pm0.000002$ & $12.98\pm0.04$ & $6.8\pm0.5$\\
              & $2.441194\pm0.000008$ & $13.54\pm0.01$ & $25.5\pm0.9$\\
              & $2.441780\pm0.000008$ & $13.44\pm0.03$ & $13.0\pm0.5$\\
              & $2.442101\pm0.000008$ & $14.03\pm0.02$ & $16.8\pm0.6$ \\
              & $2.442291\pm0.000002$ & $13.73\pm0.03$ & $8.9\pm0.3$\\
              & $2.4426484\pm0.0000008$ & $14.33\pm0.05$&$11.1\pm0.1$\\
              & $2.442886\pm0.000008$ & $14.31\pm0.01$ & $30.5\pm0.5$ \\
              & $2.443153\pm0.000001$ & $13.42\pm0.01$ & $3.5\pm0.2$\\
              & $2.443519\pm0.000001$ & $13.82\pm0.01$ & $10.0\pm0.1$\\
              & $2.443752\pm0.000008$ & $12.51\pm0.04$ & $4.2\pm0.6$\\
              & $2.444066\pm0.000003$ & $13.08\pm0.01$ & $14.7\pm0.7$ \\
              & $ 2.444396\pm0.000008$ & $12.57\pm0.03$ & 10 (f) \\
              & 2.444680 (f) & $12.20\pm0.03$ &  8 (f) \\

\hline
\end{tabular}
\end{table}

\centerline{\bf System at \zabs~$\sim 2.4426$ towards HE1158-1843}

\noindent
A $\sim 350$ \kms\ wide system is observed at an average velocity of 
$\sim 900$ \kms\ from the systemic redshift of QSO
HE1158-1843 with associated \CIII, \CIV, \NV\ and strong \OVI\ absorption
lines, as shown in Fig.~\ref{he1158_z244}. 
\NIII, \SiIII\ and \SIII\ are blended if present,
while \SiIV\ and \SIV\ are not observed.   
In Fig.~\ref{part_cov} we show that the intensity ratio of the
lines of the \CIV\ doublet is not 2 as expected, this could be due to
an incomplete coverage of the continuum source by the absorber.        
The unnatural ratio of the \NV\ doublet lines and  the disagreement
between the \HI\ \Lya\ line and the \Lyb, \Lyg\ line intensities
confirm the hypothesis of partial coverage. 
In order to measure column densities for the observed ions we
artificially fixed the zero level at the bottom of the stronger lines 
of the \CIV\ and \NV\ doublets and carried out the fitting.   
We followed the same procedure for the \Lya\ for which we adopted the
redshifts of the main \CIV\ components.  The column density obtained
for the central component of \Lya\ is consistent with the velocity
profiles of \Lyb\ and \Lyg. 
They are reported in Table~\ref{cd_he1158}. 

The upper limits on the column densities for the undetected  \SiIV\
and \SIV\ transitions  are $\log N($\SiIV$) < 11.35$ and $\log N($\SIV$) <
12.9$ where we have adopted the Doppler parameter of \CIV\ for the 
component at \zabs~$\simeq 2.4426$.  

\noindent
The partial coverage effect suggests that the size of the absorbing
cloud is comparable with that of the continuum source, in the
hypothesis of spherical geometry it is possible to derive a lower
limit to the absorber total density \citep{ppjeal94}: 

\begin{equation}
\label{part_cov_lim}
n > 5\ {\rm cm}^{-3} \left(\frac{N({\rm HI})}{10^{14.2}\ {\rm
    cm}^{-2}}\right) \left(\frac{{\rm HI/H}}{10^{-5}}\right)^{-1}
    \left(\frac{r_{\rm cont}}{1\ {\rm pc}}\right)^{-1}
\end{equation}

\noindent
where we have adopted $r_{\rm cont}= 1$ pc as the extension of the ionizing
source considering that both the broad line region and the continuum
emitting region contribute to the ionizing flux. 

\noindent
We computed the chemical abundances only for the strongest component at
\zabs~$=2.4426$ for which the column densities of both \HI\ and metals
where best determined.

\vskip 12pt

\begin{figure}
\includegraphics[width=8cm,height=9cm]{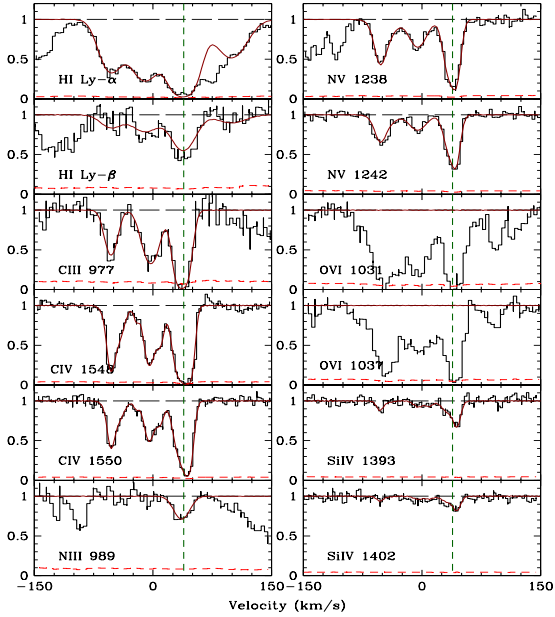}
\caption{Ionic transition lines observed at redshift $z=2.5695$ (marked by
  the vertical dotted line) in the spectrum of QSO Q2343+1232. The result
  of the best fitting for the analysed absorptions is overplotted on
  the spectrum. The short-dashed
  line represents the noise}
\label{q2343_z257}
\end{figure}

\begin{table}
\caption{Parameters obtained fitting the absorption system
  at \zabs~=2.5697 in the spectrum of Q2343+1232}
\label{cd_q2343}
\begin{tabular}{llcc}
\hline
Ion & Redshift & log N & b  \\
& & (cm$^{-2}$) & (\kms) \\
\hline 
\HI$\ldots$  & 2.568871(f) & $13.42 \pm 0.01$ & $17.7\pm 0.5$ \\
             & $2.569381\pm0.000008$ & $13.65\pm0.02$ & $22\pm1$ \\
             & $2.569957\pm0.000008$ & $13.99\pm0.01$ & $19.5\pm0.6$ \\
             & $2.570681\pm0.000008$ & $13.34\pm0.02$ & $24\pm1$ \\  
\CIII$\ldots$& $2.568868\pm0.000008$ & $12.89\pm0.04$ & $10\pm1$ \\
             & $2.56947 \pm0.000008$ & $13.15\pm0.03$ & $14\pm1$ \\
             & $2.569936\pm0.000004$ & $13.70\pm0.05$ & 9(f) \\
\CIV$\ldots$ & $2.568844\pm0.000003$ & $13.19\pm0.05$ & $4.2\pm0.5$ \\	   
             & $2.568909\pm0.000006$ & $13.31\pm0.04$ & $11.6\pm0.5$ \\	   
             & $2.569232\pm0.000006$ & $12.36\pm0.06$ & 3(f) \\		   
             & $2.569424\pm0.000005$ & $13.43\pm0.03$ & $7.6\pm0.6$ \\	   
             & $2.569587\pm0.000006$ & $12.99\pm0.07$ & $5.4\pm0.8$ \\	   
             & $2.56990\pm0.00001$ & $13.78\pm0.08$ & $9.9\pm0.6$ \\	   
             & $2.570007\pm0.000006$ & $13.90\pm0.07$ & $6.0\pm0.4$ \\     
\NIII$\ldots$& $2.56994\pm0.00002$ & $13.46\pm0.05$ & $13\pm2$ \\ 
\NV$\ldots$  & 2.568869(f) & $13.20\pm0.05$ & $6.9\pm0.8$ \\
             & 2.568934(f) & $13.28\pm0.05$ & $15\pm1$ \\
             & $2.569434\pm0.000008$ & $13.34\pm0.02$ & $14\pm1$ \\
             & $2.56993\pm0.00002$ & $13.75\pm0.09$ & $9.2\pm0.6$ \\
             & $2.570023\pm0.000008$ & $13.46\pm0.16$ & $4\pm2$ \\
\SiIV$\ldots$ & $2.568852\pm0.000003$ & $11.73\pm0.09$ & $4.2\pm0.5$ \\	   
	      & $2.569432\pm0.000005$ & $11.8\pm0.1$ & $7.6\pm0.6$ \\	   
	      & $2.569596\pm0.000007$ & $11.6\pm0.1$ & $5.4\pm0.8$ \\	   
	      & $2.56990\pm0.00001$ & $12.29\pm0.08$ & $11\pm2$ \\	   
 	      & $2.570019\pm0.000007$ & $12.24\pm0.07$ & $4.8\pm0.8$ \\
\hline
\end{tabular}
\end{table}

\centerline{\bf System at \zabs~$=2.5697$ towards Q2343+1232}

\noindent
A system with three main components is observed in the spectrum of
Q2343+1232, redshifted by $\sim 1700$ \kms\ with respect to the QSO
emission lines.    
It shows absorptions due to \CIII, \CIV, \NV\ and a weak \SiIV. \NIII\ is
detected only at the redshift of the strongest component. The upper
limits on the other two main components obtained adopting the Doppler
parameter of \CIII\ are: $\log N($\NIII$)< 13$ at \zabs~$= 2.568876$
and $\log N($\NIII$)< 13.1$ at \zabs~$=2.569432$.  
The \HI\ column density was obtained by the simultaneous fitting of the
\Lya\ and \Lyb\ absorption complexes which look free from blending. 
The detected ionic transitions are shown in Fig.~\ref{q2343_z257}
and the measured column densities are reported in Table~\ref{cd_q2343}.

\vskip 12pt

\begin{figure}
\includegraphics[width=8cm,height=8cm]{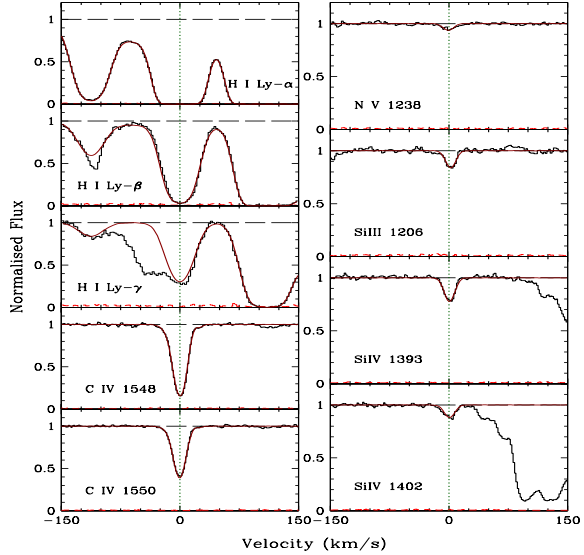}
\caption{Ionic transition lines of the absorption system at redshift
  $z=2.63618$ (marked by the vertical dotted line) in the spectrum of QSO
  Q0453-423. The result of the best fitting for the analysed
  absorptions is overplotted on the spectrum. The short dashed line
  represents the noise}  
\label{q0453_z263} 
\end{figure}

\begin{table}
\caption{Parameters obtained fitting the absorption system at
  \zabs~=2.63618 in the spectrum of Q0453-423}
\label{cd_q0453}
\begin{tabular}{llcc}
\hline
Ion & Redshift & log N & b  \\
& & (cm$^{-2}$) & (\kms) \\
\hline 
\HI$\ldots$& $2.6361797\pm0.0000008$ & $14.76\pm0.01$ & $20.0\pm0.1$ \\
\CIV$\ldots$ & $2.636071\pm0.00002$ & $12.7\pm0.1$ & $12.4\pm0.8$ \\
             & $2.636187\pm0.000001$ & $13.60\pm0.01$ & $8.7\pm0.1$ \\
             & $2.636484\pm0.000006$ & $11.60\pm0.04$ & 3(f) \\ 
\NV$\ldots$  & $2.636167\pm0.000008$ & $12.43\pm0.03$ & $12\pm1$ \\
\SiIII$\ldots$ & $2.636205\pm0.000003$ & $11.72\pm0.02$ & $7.6\pm0.4$ \\
\SiIV$\ldots$ & $2.636187\pm0.000001$ & $12.34\pm0.01$ & $7.9\pm0.2$ \\
\hline
\end{tabular}
\end{table}

\centerline{\bf{System at \zabs~$=2.63618$ towards Q0453-423}}

\noindent
A \CIV\ doublet  is detected at this redshift (\zem~$-$~\zabs$\sim 2040$
\kms) 
showing a very weak associated \NV\ $\lambda\,1238$ absorption and also
weak \SiIII\ and \SiIV\ (see Fig.~\ref{q0453_z263}).  
\OVI\ and \CIII\ are possibly present although they are blended. 
The upper limit on the \NIII\ and \CIII\ column densities adopting the
redshift and Doppler parameter of \SiIII\ are $\log N($\NIII$)<12.8$
and $\log N($\CIII$)<13.55$. 
The \HI\ \Lya\ and \Lyb\ lines were used to determine the column
density: $\log N($\HI$)\simeq 14.763\pm 0.004$ which is consistent
with the blended \Lyg\ absorption. All the measured column
densities are reported in Table~\ref{cd_q0453}. 

\vskip 12pt

\begin{figure}
\includegraphics[width=8cm,height=8cm]{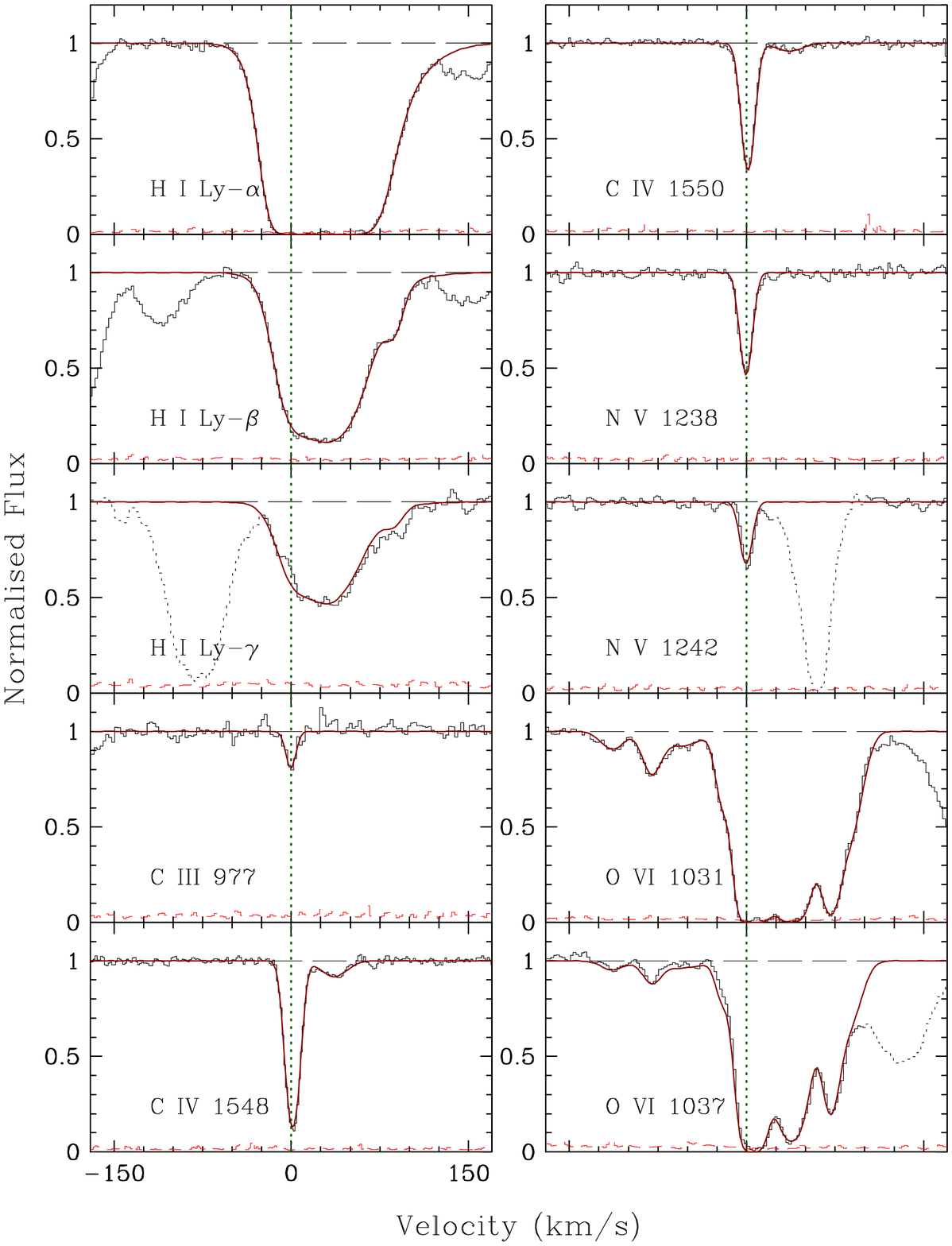}
\caption{Ionic transition lines of the absorption system at redshift
  $z=2.70897$ (marked by the vertical dotted line) in the spectrum of QSO
  PKS0329-255. The result of the best fitting for the analysed
  absorptions is overplotted on the spectrum. The short dashed line
  represents the noise}  
\label{pks0329_z270} 
\end{figure}

\begin{table}
\caption{Parameters obtained fitting the absorption system at
  \zabs~=2.7089 in the spectrum of PKS0329-255}
\label{cd_pks0329}
\begin{tabular}{llcc}
\hline
Ion & Redshift & log N & b  \\
& & (cm$^{-2}$) & (\kms) \\
\hline 
\HI$\ldots$& $2.708917\pm0.000008$ & $14.01\pm0.05$ & $15.9\pm0.8$ \\
           & $2.709304\pm0.000008$ & $14.75\pm0.02$ & $32.6\pm0.6$ \\
           & $2.7096\pm0.0001$ & $13.6\pm0.15$ &  $53\pm3.0$ \\
\CIII$\ldots$& $2.70891\pm0.00001$ & $12.07\pm0.05$ &  $4\pm2$ \\
\CIV$\ldots$ & $2.708939\pm$ & $13.589\pm0.005$ &  $6.0 \pm0.1$ \\
             & $2.709371\pm0.000006$ & $12.47 \pm0.03$ &  $15\pm1$ \\
\NV$\ldots$  & $2.708913\pm0.000002$ &  $13.29\pm0.01$ &  $5.5\pm0.2$ \\ 
\OVI$\ldots$ & $2.70751\pm0.00001$ & $12.81\pm0.04$ &  $13\pm2$ \\
             & $2.70793\pm0.00001$ & $13.09\pm0.08$ &  $10\pm1$ \\
             & $2.70824\pm0.00007$ & $12.9\pm0.1$ &  $23\pm8$ \\
             & $2.70867\pm0.00001$ & $13.25\pm0.08$ & $7\pm1$ \\
             & $2.708977\pm0.000003$ & $15.1\pm0.1$ & $9.5\pm0.8$ \\
             & $2.709390\pm0.000004$ & $14.66\pm0.01$ & $15.6\pm0.6$ \\
             & $2.709800\pm0.000004$ & $14.12\pm0.04$ & $8.0\pm0.5$ \\
             & $2.70998\pm0.00002$ & $13.79\pm0.07$ & $14\pm1$ \\
\hline
\end{tabular}
\end{table}

\centerline{\bf{System at \zabs~$=2.708968$ towards PKS0329-255}}

\noindent
This system is redshifted by $\sim 1940$ \kms\ from the QSO emission
lines. It shows optically thin absorptions
due to \CIV, \CIII\ and \NV. \OVI\ is also  detected, 
it is saturated and with at least two other components besides
the one in common with the other ions (see Fig.~\ref{pks0329_z270}).
The measured column densities for the observed transitions are reported in
Table~\ref{cd_pks0329}.  
The upper limits on the column densities of \NIII\ and \SiIV\ obtained
adopting the Doppler parameter of \CIII\ and \CIV, respectively are: $\log
N($\NIII$) < 12.5$ and $\log N($\SiIV$) < 11.5$.  
The \HI\ column density corresponding to the metal absorption is well
constrained by the simultaneous fitting of the \Lya, \Lyb\ and \Lyg\  
transitions which gives: $\log N($\HI$)\simeq 14.01 \pm 0.05$. 

\section{The photoionisation model}

In order to obtain elemental abundances from the observed 
ionic column densities it is necessary to compute ionisation 
corrections. 
To this purpose we used the code Cloudy \citep{ferland03} adopting  
as an ionising spectrum a typical QSO spectrum derived from
\citet{cristiani:vio} and extrapolated in the region shortward of the
\Lya\ emission with a power law  $f(\nu) \propto \nu^{-0.9}$
corresponding to the continuum slope observed redward of the \Lya\
emission.  
At energies higher than the Lyman limit, we adopted a power law
$f(\nu) \propto \nu^{-1.60}$ following \citet{hamanneal01}. This is
consistent with the best observations of luminous QSOs and with the
slope of the UV background at high energies. 
We verified that results did not change using a slightly different
composite spectrum obtained from the FIRST bright QSO survey by 
\citet{FBQS}.  The absorption in the range $912 < \lambda_{\rm rest} <
1216$ \AA\ was corrected adopting the power law fitting the 
spectrum at larger wavelengths: $f(\nu) \propto \nu^{-0.46}$ and for
$\lambda_{\rm rest} < 912$ \AA\ we used $f(\nu) \propto \nu^{-1.60}$ as
before.   

We assumed that components of different transitions which are at the
same velocity arise in the same gas and we tried to reproduce with
Cloudy all the ionic column densities observed at the same redshift 
with a single-region model. 

As a zero-order approach we computed for each system a grid of Cloudy 
models for a cloud with the measured \HI\ column density, solar
metallicity and relative abundances, varying the ionisation parameter,
which is defined  as: 

\begin{equation}
U \equiv \frac{1}{4 \pi\, c\, r^2_0 n_{\rm H}}\int_{\nu LL}^{\infty} 
\frac{L_{\nu}}{h\nu}d\nu, 
\end{equation}

\noindent
where $L_{\nu}$ is the luminosity density of the QSO spectrum, $c$ is the
speed of light, $r_0$ is the separation between the QSO  and the
illuminated face of the  cloud and $n_{\rm H}$ (cm$^{-3}$) is the total
hydrogen density (\mbox{H\,{\sc ii}} + \HI\ + H$_2$). 

In every studied system it was not possible to reproduce all the observed
ionic column densities with a single $U$ value, even allowing for a
vertical shift to account for a non-solar metallicity. 
This was true also when considering groups of ions of similar ionisation
state.
Then, this discrepancy cannot be ascribed (only) to the fact that we are
treating as a single region a multi-phase gas, and to be solved it requires
the introduction of non-solar relative abundances.   

To derive the relative abundances and reproduce the observed set of
column densities we followed three main steps. 
\par\noindent
(1) The value of the ionisation parameter for the considered absorption
system, $U_{\rm  s}$, was fixed on the basis of the column density ratio of
two, possibly close, ionisation states of one or more elements. 
\par\noindent
(2) The column densities of all observed ions were then computed with
Cloudy for $U=U_{\rm s}$, with the measured \HI\ column density, solar
metallicity and relative abundances.
The ratios of the observed to the computed column densities were taken 
as the variations of the element abundances with respect to the solar values. 
\par\noindent
(3) As a final step, we introduced new elemental abundances in the model
computed from the variations determined in step (2). Then, instead of
fixing $U_{\rm s}$ we normalised the spectrum with the absolute magnitude
of the studied QSO and ran a grid of models varying the
value of $n_{\rm H}$ and finding the corresponding values of $r_0$ which
gave viable solutions. 

Due to the degeneracy between the total density and the radius (refer
to eq.~3), for all the reasonable values of $n_{\rm H}$ it is possible
to find a corresponding $r_0$ at which the set of column densities
matches the observed one. 
For three systems in Section~3.2 we derived physical limits on $n_{\rm H}$ 
in an independent way which consequently gave limits on the distance from
the source. 
In principle it is possible to derive a lower limit on $n_{\rm H}$,
that is an upper limit on the distance from the QSO, computing the
total density of a cloud with the observed  neutral hydrogen column
density and a characteristic dimension equal to the local Jeans
length \citep[see ][]{schaye01}.  
%
%
However, this approach, due to the low \HI\ column densities
observed, does not provide stringent limits ($r_{0,{\rm max}} \sim
10-20$ Mpc). 

We could not determine the metallicity of the gas from the abundance of
iron, since we did not observe iron in our systems neither as \FeII\ nor as
\FeIII. 
Observations of stars in our Galaxy show that [C/Fe] is consistent with
solar at least for [Fe/H]~$> -1$ \citep*{carretta00}. This is predicted by
chemical evolution models of our Galaxy where a consistent amount of carbon
is produced by low and intermediate mass stars
\citep*{timmes95,chiapp03a}, enriching the
interstellar medium on a time-scale similar to that of iron enrichment
from Type Ia SNe. 
In our analysis, we used carbon as a proxy of iron and we computed the
$\alpha$-element abundances compared to it.

\section{Results}

We investigated the star formation history in the close neighbourhood 
of six QSOs at emission redshifts between 2.1 and 2.7 deriving the 
metallicities and relative chemical abundances of six associated 
narrow absorption systems. 
The results of our calculations are summarised in Table~9 where 
errors on the abundance ratios are due mainly to the uncertainties in the 
column density determinations. 

\vskip 12pt
\noindent 
1. For the systems associated with the QSOs UM680 and UM681 we were
able to put upper limits on the total hydrogen density in an
independent way  (see  Section~3.2).  
They correspond in the Cloudy photoionisation models to lower limits
on the distance from the emitting source of $r_0 \sim$~70-120 kpc and
$r_0 \sim$~120-260 kpc  for UM680 and UM681 respectively. 
The two absorption systems fall at very close redshifts and present
similar velocity structures, evidences that suggested the presence of
a diffuse gaseous  structure possibly including the two QSOs
\citep{shav:rob,vale02}.  
The metallicity of the two systems is quite uncertain due to saturated
\HI\ \Lya\  lines. 
However, the system associated with UM680 has a [C/H]  
abundance ratio at least 5 times solar and larger than that of
UM681 which is consistent with solar. The [$\alpha$/C] is larger in
UM681 than in UM680, while the two systems have [N/C] abundances
consistent with solar. 
We speculate that the observed gas is an outflow
of UM680 which is pierced by the UM681 line of sight in an external, less 
enriched region. 
Deep imaging of the field could possibly shed some light on the nature of
these absorbers. For example by detecting \Lya-emitters at the same
redshift of the absorbers tracing a large scale structure between the two
lines of sight. 

\medskip
\noindent
2. For the absorber towards QSO HE1158-1843 it was possible to compute a lower
limit on the total density which translates into a separation from the
continuum source smaller than $\sim 40$ kpc. In this case we can assert
that we are probing the abundances in the interstellar medium of the host
galaxy. The metallicity is about twice solar and the N/C ratio is slightly
supersolar. Unfortunately  we did not detect any $\alpha$-element. 

\medskip
\noindent
3. The three components of the system in QSO Q2343+1232 show very similar 
relative abundances although the \HI\ column densities vary of a
factor of 3 among them. The metallicity and the ratio N/C are between 2
and 3 times solar and there is an indication of $\alpha$-enhancement. 

\medskip
\noindent
4. The two systems at larger redshifts are characterised by small \CIV\
equivalent widths and show different abundance patterns compared
to the other systems. In particular they have undersolar N/C ratios. 

\medskip
\noindent
5. No significant correlation is observed between the velocity
separation from the quasar and the metallicity of the system.   
This suggests that the observed separations cannot be trivially related to
the actual spatial distances between the central source and the
absorber due both to uncertainties in the emission redshift
determination and to peculiar velocities of the absorbing material. 

\vskip 12pt
As shown in Fig.~\ref{rel_ab_mod}, only one among the six studied
systems has a metallicity significantly lower than solar, $Z \sim 1/6\
Z_{\sun}$. The other 5 systems show values comparable or larger than
solar.  
We confirmed the supersolar N/C abundance ratio in those systems with
$Z\ga  Z_{\sun}$, as already found in other AALs \citep[see ][ for a 
  review]{HF99}.     
On the other hand, in our AALs we measured enhanced $\alpha$-element/C  
abundance ratios at variance with the tentative detection of 
supersolar Fe/Mg abundance ratios in broad emission line regions 
(see Section~7) but in agreement with abundances measured in
elliptical galaxies.  
This suggests that we are sampling regions where SNe Ia did not yet
have the time to enrich the gaseous medium, implying that the bulk of star
formation started less than $\sim 1$ Gyr before.    

\begin{figure}
\includegraphics[width=8cm,height=8.5cm]{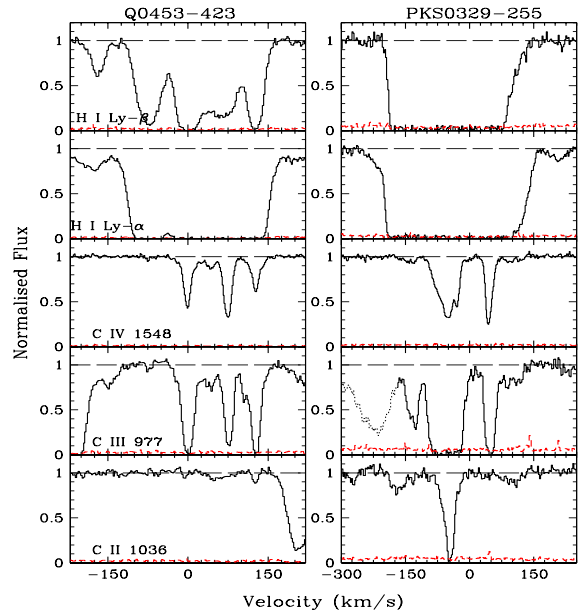}
\caption{Observed intervening systems at \zabs~$=2.4426$ in the spectrum of
Q0453-423 (left panels) and at \zabs~$=2.4552$ in the spectrum of
PKS0329-255 (right panels)}
\label{intervening}
\end{figure}

Due to the lack of constraints on the total density of the absorbers
we cannot reliably establish how close they are to the continuum
source in all but one case.  However, adopting a density typical of the
interstellar medium, $n_{\rm H}=10$ cm$^{-3}$, we obtained separations 
of the order of $100$ kpc. To reach such a distance in $0.5 -1 \times
10^8$ yrs, the average lifetime of a QSO, the gas should have
travelled at a velocity of $\sim 1000-2000$ \kms, which is indeed of
the order of the $\Delta v$  measured for our systems.  

Another way to test the hypothesis that these systems are under the
strong influence of the associated QSO and probe chemical abundances in
the QSO vicinity is to compare them with intervening
systems with similar \CIV\ equivalent width. 
In Fig.~\ref{intervening} we show two examples of intevening systems
at velocity separations of $\sim 18400$ and $19300$ \kms\ from
the QSOs Q0453-423 and PKS0329-255 respectively.     
It is apparent that the ionisation pattern is completely different
from the one observed in AALs, in particular the \CIII\ transition
is much stronger than the \CIV\ $\lambda\,1548$ line while in the
studied AALs the \CIII/\CIV\ column density ratio is always lower than
one. 
This suggests that most optically thin intervening systems are ionised
by the diffuse UV background while associated systems receive the
direct emission from the QSO. In particular, the ionisation 
energy of the \CIII~$\rightarrow$~ \CIV\ transition is close to a break
present in the UV background spectrum at the He ionisation energy which is
absent in the unabsorbed QSO spectra.     

The analysis of abundances in intervening systems is a difficult task
because in many cases it is not possible to derive with confidence the
\HI\ column density and, in general, there is an ambiguity on 
which kind of sources is ionising the gas. 
However, the vast majority of the studies on intervening Lyman limit
and optically thin metal absorption systems carried out up to now
found undersolar metallicities   
\citep[e.g. ][]{berg94,kohlereal99,proch:burl,dodo:ppj01}. 
Finally, the damped \Lya\ systems for which the determination of
abundances is very precise due to negligible ionisation corrections, 
are characterised by an average metallicity lower than 1/10 solar at
$z\geq 1$.

We conclude that the associated systems show indeed a peculiar chemical
abundance pattern and in the following section we try to frame our
results into a model for the chemical evolution of large elliptical
galaxies.

\section{Comparison with model predictions}

We compared the present results on abundances with theoretical
predictions based on a physical model for the co-evolution of QSO
and host galaxy systems \citep{romano02,granato04}.  
In order to illustrate the effects of the star-formation rate (SFR)
history on abundances, we considered two cases for the gas
distribution inside the virialized DM  halo. 
The first one, case A, assumes that the gas closely follows the DM 
profile with no clumpiness, while in the second case, B, we assumed 
that after virilization on the average the gas follows the DM profile,
but we introduced a clumping factor. As expected, in case A it is difficult
to get a rapid star formation in DM haloes with $M_{\rm halo}\geq 
1.5 \times 10^{13}$ M$_{\sun}$ before the QSO shines \citep[see Fig.~10 in
][]{romano02}, while in case B we can form stars very rapidly 
even in larger haloes \citep[see][]{granato04}. 

Both cases consider a single-zone galaxy and a double power-law IMF,
i.e. $\Phi($M$) \propto~$M$^{-0.4}$ for M~$\le 1$~M$_{\sun}$ and
$\Phi($M$) \propto~$M$^{-1.25}$ for 1 M$_{\sun} \le$~M~$\le
100$~M$_{\sun}$ (in a notation where the Salpeter index would be
1.35). 
We also used for both cases the chemical yields adopted by \citet[][
  their model 7]{chiapp03a}.  
This yield set was chosen because it proved to give the best agreement with
CNO osbervations for the Milky Way, the M101 spiral galaxy, DLAs and dwarf
irregular galaxies.       
  
\begin{figure}
\includegraphics[width=8cm,height=8cm]{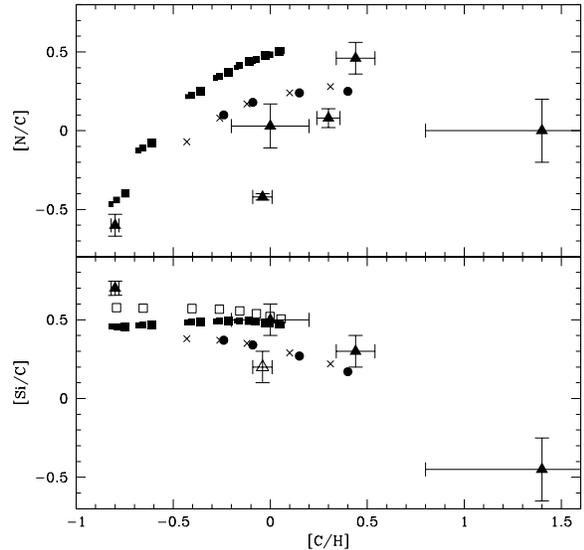}
\caption{[N/C] and [Si/C] abundance ratios vs. [C/H] obtained for the
  analysed associated absorption systems and for the considered
  chemical evolution model. Solid triangles with error bars are our
  data. The abundances of the three components of the system towards
  Q2343+1232 have been averaged to a single value. 
  The empty triangle in the bottom
  panel is the [O/C] abundance ratio for the system towards PKS0329-255. 
  {\sl Upper panel:} solid dots are the predictions of case B based on 
  \citet{granato04} for a DM halo mass of $3\times 10^{13}$
  M$_{\sun}$, $z_{\rm QSO} = 2.1$ and $z_{\rm vir}=2.4$, 2.5, 2.7 and 3 in
  order of increasing [C/H] abundance. The crosses are the predictions for
  a DM mass of  $5\times 10^{13}$ M$_{\sun}$, $z_{\rm QSO} = 2.55$ and
  $z_{\rm vir}=2.8$, 2.9, 3, 3.2, 3.5. Solid squares are the results
  of case A based on \citet{romano02} for a DM halo of M$_{\rm
  halo} =1.37 \times 10^{13}$ M$_{\sun}$. Increasing sizes represent 
  virialization redshifts $z_{\rm vir}= 2.5, 3$ and 4, the different groups
  of squares are the abundances at time-steps of 0.1 Gyrs from $z_{\rm
  vir}$ to $z_{\rm QSO}$. {\sl Lower panel:} Solid dots, crosses and solid
  squares are the same as above.  Empty squares represent the predictions
  by \citet{romano02} for the abundance of [O/C] starting at $z_{\rm
  vir}=3$}
\label{rel_ab_mod}
\end{figure}

For case A we simulated a halo of mass M$_{\rm halo} =1.37 \times
  10^{13}$ M$_{\sun}$ \citep[model 1d in ][]{romano02}. 
We followed the evolution of gas abundances from the beginning of star
formation to the shining of the QSO at time-steps of 0.1 Gyrs and for
different virialization redshifts. 
The model predictions reported in Fig.~\ref{rel_ab_mod} reproduce the
correlation [N/C] vs. [C/H] and the $\alpha$-enhancement, but the 
relation between [Si/C] and [C/H] is almost flat. 

The predictions for case B \citep[see ][]{granato04} refer to the
chemical evolution of two DM haloes of masses M$_{\rm halo} = 3 \times
10^{13}$ and $5 \times 10^{13}$ M$_{\sun}$ corresponding roughly to the
absolute magnitudes $M_{\rm B} = -27.5$ and $-29.5$ in the hypothesis that
the QSOs are shining at the Eddington luminosity and applying eq. 6 by 
\citet{ferrarese02}: 

\begin{equation}
\frac{M_{\rm BH}}{10^8\ M_{\sun}} \sim 0.10 \left( \frac{M_{\rm
    halo}}{10^{12}\ M_{\sun}}\right)^{1.65}.
\end{equation}  

We explored virialization redshifts at time intervals of $\sim
0.5$ and 1 Gyr from the QSO emission redhifts in our sample. 
As shown in Fig.~\ref{rel_ab_mod}, the predictions cover the range of
[C/H] where most of the observations lie. They reproduce both the
observed correlation between [N/C] and [C/H] and the anticorrelation
in [Si/C] vs. [C/H]. 

The differences between the predictions of case A and B are small. 
They are mostly due to the higher SFR that can be attained in case B
before the QSO shining, due to the clumping factor that shortens the
cooling time of the gas. 
It should be noticed also that this is a single-zone model,
i.e. the results are averaged over the whole physical dimension of the
galaxy, while a metallicity gradient is observed in elliptical galaxies. 

In conclusion, the high level of chemical enrichment and the
$\alpha$-enhancement observed in the QSO environments indicates that
the massive elliptical galaxies hosting QSOs must have formed the bulk
of their stellar population on short time-scales at high redshifts. 

\section{Alternative measurements of chemical abundances in
  the vicinity of QSOs}

Our results are in general agreement with QSO chemical abundances
determined with other methods. 

Broad emission lines (BELs) observed in QSO spectra are the most
commonly used diagnostics to this purpose \citep[see
  e.g. ][]{david:netzer,HF99}.  
BELs are known to originate in photoionised gas within $\sim
1$ pc of the central continuum source. 
It has become the norm in BEL studies to take the N/O abundance ratio 
as a tracer of O/H, that is of $Z$. 
The prominent metal lines, such as \CIV\ $\lambda\,1549$, relative to \Lya\  
are not sensitive to the overall metallicity for $Z> 0.1Z_{\sun}$
\citep{HF99}. 
On the other hand, observations in \mbox{H\,{\sc ii}} regions indicate
that the N/O~$\propto$~O/H relation is valid for metallicities above
$\sim 1/3$ to $\sim 1/2$ solar
\citep{shields76,pagel:edm81,vanzee98,izot:thuan99,pettini02}.  
This abundance behaviour is attributed to ``secondary'' N production,
whereby N is synthesized from existing C and O via CNO burning in stars.  

Calculations of BEL metallicities in large samples of QSOs spanning the
redshift range $0 \le z_{\rm em} \le 5$ found typically solar or supersolar
metallicities across the entire redshift range and no evidence of a  
decrease at the highest redshifts 
\citep[][]{HF93,hamanneal02,warnereal02,dietreal03a,warnereal03}. 
Precise estimates are difficult because metallicities derived
from the \NV\ lines (most notably \NV/\mbox{He\,{\sc ii}}) are typically
$\sim 30$ \% to a factor of $\sim 2$ larger than estimates from the
intercombination ratios (e.g. \NIII$]$/\CIII$]$). 
The reason for this discrepancy is not clear and also the absolute
uncertainties are not easily quantified because they depend on the assumed
theoretical models.  

The determination of the relative abundance of iron versus $\alpha$-elements,  
which is taken as an indicator of the time elapsed from the beginning of  
the last star formation episode, relies for QSO BELs on the ratio 
\FeII(UV)/\MgII\ $\lambda\,2798$, where ``\FeII(UV)'' indicates a broad
blend of many \FeII\ lines between roughly 2000 and 3000 \AA\ which is very
hard to measure.  
The tentative measurements carried out at low and intermediate redshifts 
indicate Fe/Mg a factor of $\sim 3$ above solar \citep{wills85} suggesting 
that SN Ia already contributed to the gas enrichment.
Recently, new measurements of the \FeII/\MgII\ ratio were obtained for high
and very high ($z\sim 6$) redshift QSOs showing no clear evolution with
time \citep{freudeal03,dietreal03b}. 
However the consequences of these results on the early star formation history 
will be clearer only when the theoretical relationship 
between the observed \FeII/\MgII\ emission ratio and the Fe/Mg abundance will 
be assessed \citep{vernereal99,vernereal03,sigut:prad03}.

\vskip 12pt

Also ``broad'' absorption lines (BALs) have been used to investigate
the chemical and physical properties of the gas associated with QSOs. 
Indeed, BALs are believed to arise in material ejected by the
QSO but still located very close to the central regions. 
The drawback in the use of these systems is that broad profiles blend
together all the important doublets and do not allow a 
reliable estimate of the column densities, in particular in the case of
partial coverage of the source. 
However, the  numerous studies on BALs indicate metallicities near or above 
the solar value \citep[e.g.][]{korista96,hamann98,arav01,sria:ppj01,gupta03}.

\section{Conclusions}

Up to now the main approach to study the chemical abundances in QSO
environments has been the analysis of BELs observed in their spectra. 
Metallicities determined from BELs are consistent with solar or slightly
supersolar values without a significant evolution in redshift. Other
elemental abundances are very difficult to measure, in particular
determinations of the ratio $\alpha$/Fe are very uncertain. 

Associated narrow absorptions are complementary probes of the physical
status of QSO-elliptical systems with respect to BELs. In general, 
they can be due to gas belonging to the interstellar medium of the
galaxy, outflowing under the effect of the QSO or re-infalling on the
QSO itself. Furthermore, it is more straightforward to derive chemical
abundances from absorption lines than from emission lines.  
We need only to determine and apply the proper ionisation
corrections to convert the measured ionic column densities into
relative abundances.  

In this paper, we selected six narrow absorption systems lying within
5000 \kms\ from a QSO emission redshift and determined the abundances
of C, N and $\alpha$-elements in the gas they originate from.  
We used high resolution, high signal-to-noise UVES QSO spectra and
applied a procedure based on the photoionisation code Cloudy to
compute the chemical abundances starting from the measured column
densities.  
    
We found that all systems but one in our sample have metallicities
(measured by carbon) consistent with or larger than solar. 
We found also a possible correlation of [N/C] and an anticorrelation
of [Si/C] with [C/H] with supersolar values of [Si/C].   
These results are suggestive of rapid enrichment due to a short star
formation burst, of duration $t_{\rm burst}\sim$ 1 Gyr (see
Section~5).  
Since the very high luminosity QSOs in our sample should have
$M_{\rm BH}\geq 10^9$ $M_{\sun}$ , assuming $M_{\rm sph}/M_{\rm
  BH}\sim 1000$ \citep{mclure:dunlop} we expect SFR$\geq 1000$
$M_{\sun}$ yr$^{-1}$ in their hosts.
 
The predictions of the model of chemical evolution for a spheroidal
galaxy where the star formation depends on stellar and QSO feedback
are in good agreement with the observations. In particular, the
agreement improves when taking into account a clumping factor
\citep{granato04}, which allows the gas to be efficiently converted
into stars also in very massive dark haloes with  SFR$\geq 1000$
$M_{\sun}$ yr$^{-1}$.  

In this way, narrow associated QSO absorption systems proved to be 
extremely useful in the study of the QSO environment, in particular 
when there is evidence of their intrinsicness.  
They can be used as estimators of the chemical abundances in high redshift
spheroidal galaxies which are not easily determined otherwise. 
The probed gas will probably  be ejected from the galaxy due to the QSO
feedback, thus we are also observing a potential source of enrichment
of the intergalactic medium at high redshift.  

In order to obtain a deeper insight in the evolution of QSO
host-galaxies and environments it is essential to enlarge the data
sample. In particular, obtaining high signal-to-noise spectra in the
UV to reliably measure the doubly-ionised lines of C and N and
increasing the redshift range especially at large values. Indeed, the
five $z\sim 4$ AALs analised up to now \citep{savaglioeal97} seems to
indicate a slightly 
lower average metallicity, [C/H]~$\sim-0.5$, than for the bulk of the
sample at redshift $z \sim 2 - 2.5$.  More data will be fundamental to 
verify the observed correlations and to constrain the  
predictions of theoretical models.

\section*{Acknowledgements}
We are grateful to the referee for her comments and suggestions which
greatly improved this paper. 
V.D. would like to thank MIUR/COFIN for financial support. 
This research was partially funded by ASI contracts I/R/35/00, I/R/088/02. 
Part of the work was supported by the European Community Research and
Training Network ``Physics of the Intergalactic Medium''.
We thank the ESO support astronomers who have performed some of
the observations in service mode. 

\bibliographystyle{aa} 
\bibliography{aamnem99,myref} 

\newpage

\begin{figure*}
\includegraphics[angle=90]{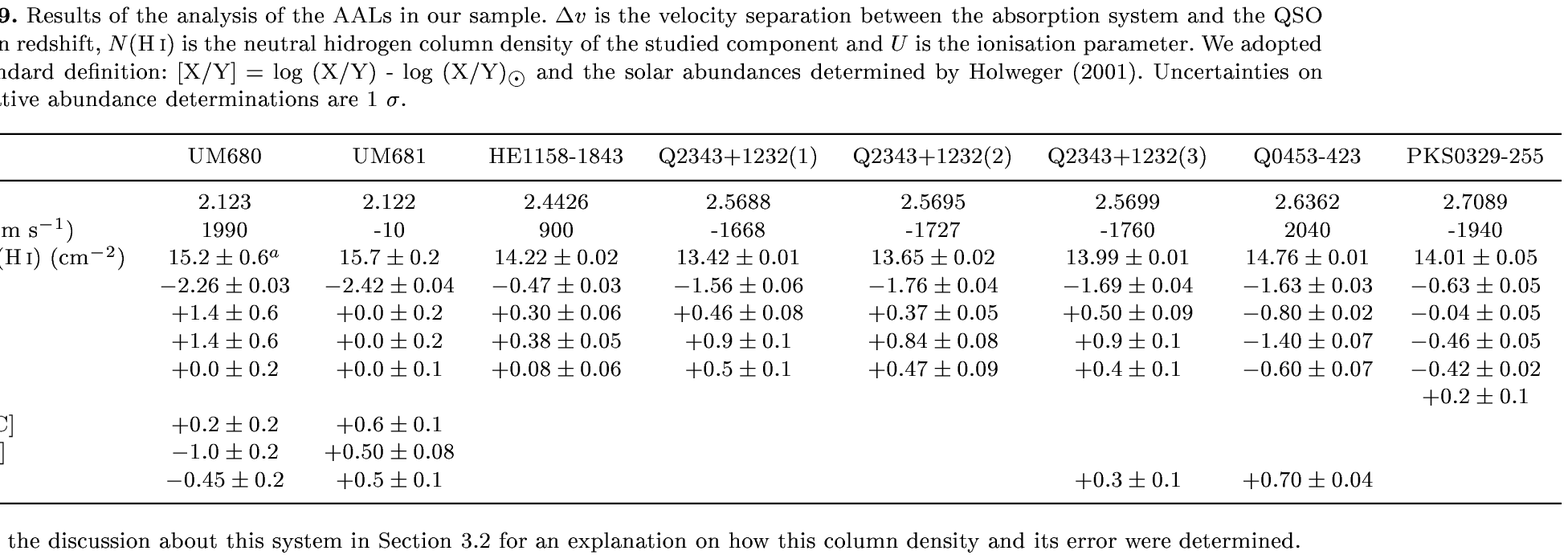}
\label{summary}
\end{figure*}

\label{lastpage}

\end{document}